\documentclass[preprints,article,accept,moreauthors,pdftex]{Definitions/mdpi} 

\usepackage{epigraph}

\firstpage{1} 
\pubvolume{xx}
\issuenum{1}
\articlenumber{5}
\pubyear{2019}
\copyrightyear{2019}
\history{}

\pdfoutput=1

\usepackage{color}
\usepackage{verbatim}
\usepackage{xcolor}
\usepackage{subcaption}
\usepackage{threeparttable}
\usepackage{makecell}

\Title{Whole-brain models to explore altered states of consciousness from the bottom up}

\Author{Rodrigo Cofr\'{e} $^{1,*}$\orcidA{}, Rubén Herzog $^{2}$\orcidB{}, Pedro A.M. Mediano $^{3}$\orcidC{}, Juan Piccinini $^{4,5}$, Fernando~E.~Rosas~$^{6,7,8}$\orcidF{}, Yonatan Sanz Perl $^{4,9}$\orcidD{}, Enzo Tagliazucchi $^{4,5}$\orcidE{}}

\AuthorNames{Rodrigo Cofre, Rubén Herzog, Pedro A.M. Mediano, Juan Piccinini, Fernando E. Rosas, Yonatan Sanz Perl, and Enzo Tagliazucchi}

\address{%
$^{1}$ \quad CIMFAV-Ingemat, Facultad de Ingeniería 2340000, Universidad de Valparaíso, Valparaíso, Chile\\
$^{2}$ \quad Centro Interdisciplinario de Neurociencia de Valparaíso 2360103, Universidad de Valparaíso, Valparaíso, Chile\\
$^{3}$ \quad Department of Psychology, University of Cambridge, Cambridge CB2 3EB, UK\\
$^{4}$ \quad National Scientific and Technical Research Council, Buenos Aires, Argentina\\
$^{5}$ \quad Buenos Aires Physics Institute and Physics Department, University of Buenos Aires, Buenos Aires, Argentina\\
$^{6}$ \quad Centre for Psychedelic Research, Department of Brain Science, Imperial College London, London SW7 2DD, UK \\
$^{7}$ \quad Data Science Institute, Imperial College London, London SW7 2AZ, UK\\
$^{8}$ \quad Centre for Complexity Science, Imperial College London, London SW7 2AZ, UK\\
$^{9}$ \quad Universidad de San Andr\'es, Buenos Aires, Argentina
}

\corres{Correspondence: rodrigo.cofre@uv.cl}

\abstract{The scope of human consciousness includes states departing from what most of us experience as ordinary wakefulness. These \emph{altered states of consciousness} constitute a prime opportunity to study how global changes in brain activity relate to different varieties of subjective experience. We consider the problem of explaining how global signatures of altered consciousness arise from the interplay between large-scale connectivity and local dynamical rules that can be traced to known properties of neural tissue. For this purpose, we advocate a research program aimed at bridging the gap between bottom-up generative models of whole-brain activity and the top-down signatures proposed by theories of consciousness. Throughout this paper, we define altered states of consciousness, discuss relevant signatures of consciousness observed in brain activity, and introduce whole-brain models to explore the mechanisms of altered consciousness from the bottom-up. We discuss the potential of our proposal in view of the current state of the art, give specific examples of how this research agenda might play out, and emphasise how a systematic investigation of altered states of consciousness via bottom-up modelling may help us better understand the biophysical, informational, and dynamical underpinnings of consciousness.}


\begin{document}

\section{Introduction}

Consciousness has been for centuries a puzzle beyond the scope of natural science; however, the significant progress seen during the last 30 years of research suggests that a rigorous scientific understanding of consciousness is possible ~\cite{ledoux2020little, seth2018consciousness, Overgaard2017}. The dawn of the modern scientific approach to consciousness can be traced back to Crick and Koch’s proposal for identifying the \emph{neural correlates of consciousness} (NCC) ~\cite{Crick1990,Crick2003}, understood as the minimal set of neural events associated with certain subjective experience. The key intuition that fuels this proposal is that careful experimentation should suffice to reveal brain events that are systematically associated with conscious (as opposed to unconscious or subliminal) perception. Needless to say, the methodological challenges associated with this idea are vast -- particularly concerning the determination of what constitutes conscious content (e.g. must content be explicitly reported, or are other less direct forms of inference equally valid? \cite{tsuchiya2015no, cohen2011consciousness}). Despite these problems,
the program put forward by Crick and Koch succeeded to jump-start contemporary consciousness research.\footnote{For recent reviews on the empirical search for NCC see Ref.~\cite{Koch2016}, for a theoretical examination of the concept of NCC see Ref.~\cite{chalmers2000neural}, and for criticism to the concept of NCC see Refs.~\cite{noe2004there, de2012correlates}.}

While the quest for the NCC aims to provide answers to \emph{where} and \emph{when} consciousness occurs in the brain, subsequent theoretical efforts have attempted to discover systematic signatures within those NCC that could reflect key mechanisms underlying the emergence of consciousness. In other words, these efforts try to answer \emph{how} consciousness emerges from the processes that give rise to the NCC \cite{seth2007models, sergent2012imaging}. Hence, theoretical models of consciousness strive to "compress" our empirical knowledge of the NCC, i.e. to provide rules that can predict \emph{when} and \emph{where} from \emph{how}. The nature of those rules, in turn, determines the kind of explanation offered by a theoretical model of consciousness. Here we consider two possible approaches: top-down and bottom-up~\cite{stinson2018mechanistic}. On the one hand, top-down approaches start by identifying high-level signatures of consciousness, and then try to narrow down low-level biophysical mechanisms compatible with those signatures. On the other hand, bottom-up approaches build from dynamical rules of elementary units (such as neurons or groups of neurons \cite{deco2008dynamic}), and attempt to provide quantitative predictions by exploring the aggregated consequences of these rules across various temporal and spatial scales. We further subdivide explanations into those addressing conscious information access (e.g. perception in different sensory modalities) and those concerning consciousness as a temporally extended state, such as wakefulness, sleep, anaesthesia, and the altered states that can be elicited by pharmacological manipulation \cite{vaitl2005psychobiology, revonsuo2009altered, overgaard2010neural,tassi2001defining, ludwig1966altered, tart1976_, bayne2007conscious}. 

Our objective is to put forward a research program for the development of bottom-up explanations for the relationship between brain activity and states of consciousness, which we claim is underrepresented both in past and current research. Theories that rely heavily on a top-down perspective risk being under-determined in the reductive sense; i.e. they could be compatible with multiple and potentially divergent lower-level biological and physical mechanisms \cite{michel2019consciousness}. 
While we do not know whether consciousness may be instantiated in other physical systems, we certainly do know that it is instantiated in the human brain, and therefore all theoretical models of consciousness should be consistent with the low-level biophysical details of the brain to be considered acceptable. In light of this potential under-determination, it is difficult to decide whether the different theories currently dominating the field are competing (in the sense of predicting mutually contradictory empirical findings) or convergent (in spite of being formulated from disparate perspectives). Without investigating theories of consciousness from the bottom-up, it could be simply too early for proposals of an \emph{experimentum crucis} to decide between candidates \cite{reardon2019rival}.

In this paper we posit that computational models can play a crucial role in determining the low-level physical and biological mechanisms fulfilling the high-level phenomenological and computational constraints of theoretical models of consciousness. The idea that consciousness is intrinsically dependent on the dynamics of neural activity is not new, and in this sense we follow the trail of pioneers such as Walter J. Freeman~\cite{freeman2007indirect}, Francisco Varela~\cite{thompson2001radical}, and Gerald Edelman~\cite{edelman2000reentry}, among others. However, our proposal reaches further than these previous attempts by building upon the technological and conceptual advances accumulated over the last decades. In particular, the widespread availability of non-invasive neuroimaging methods (fMRI, DTI, MEG) has expanded our knowledge of the functional and structural aspects of the brain, while the development of connectomics has revealed the intricate meso- and macroscopic connectivity patterns that wire cortical and subcortical structures together~\cite{sporns2005human}. Moreover, for the first time there is sufficient empirical data and computational power available to construct whole-brain models with real predictive power~\cite{deco2008dynamic, breakspear2017dynamic, ritter2013virtual}, which represents a radical improvement over past research efforts. We expect that these advances will enable us to compare the predictions of theories of consciousness by means of whole-brain computational models that can be directly contrasted with empirical results. 

In the following, we adopt and explore the consequences of this perspective. Our proposal and its justification are structured as follows. First, Section~\ref{sec:ASC} describes several examples of altered states of consciousness and briefly discusses some proposed general definitions. Next, Section~\ref{sec:data-driven_tools} introduces top-down approaches for quantifying and classifying states of consciousness solely from functional data. Then, Section~\ref{sec:wholebrainmodels} introduces the main technical ideas underlying the development of whole-brain computational models, highlighting novel results with special emphasis on those informing research on altered states of consciousness. Section~\ref{sec:proposed_agenda} discusses how computational models can contribute to overcome open challenges and conceptual difficulties, thus providing new insights into the study of altered states of consciousness. Finally, Section~\ref{sec:future} elaborates on possible future directions of research stemming from our proposal.

\section{What is an altered state of consciousness? Examples and defining features}
\label{sec:ASC}

A basic distinction is commonly drawn between phenomenal and access consciousness \cite{block1995confusion}. The first represents the subjective experience of sensory perception, emotion, thoughts, etc.; in other words, what \emph{it feels like} to perceive something, undergo a certain emotion, or engage in a certain thought process. The second represents the global availability of conscious content for cognitive functions such as speech, reasoning, and decision-making, enabling the capacity to issue first-person reports.

The term "consciousness" is also used in reference to a third concept whose definition is comparatively more elusive: that of temporally extended and qualitatively distinct modes or states of consciousness \cite{vaitl2005psychobiology, revonsuo2009altered, overgaard2010neural,tassi2001defining, ludwig1966altered, tart1976_, bayne2007conscious}. This concept is perhaps best introduced by listing examples, such as our ordinary state of conscious wakefulness, the different phases of the wake-sleep cycle, dreaming during rapid eye movement (REM) sleep, sedation and general anaesthesia, post-comatose disorders such as the unresponsive wakefulness syndrome, the acute effects of certain drugs (mainly serotonergic psychedelics and glutamatergic dissociatives), the state achieved in some contemplative traditions by means of meditation, hypnosis, and shamanic trance, among others. Following Ludwig \cite{ludwig1966altered} and Tart \cite{tart1972altered}, we refer to these as  "altered states of consciousness", adopting this term to emphasise their dissimilarity to ordinary conscious wakefulness.

Let us describe commonalities shared by altered states of consciousness, which point towards a potential general definition. First, altered states of consciousness are temporally extended and typically (but not always) reversible. Second, they are not defined by the presence of specific subjective experiences, but instead by general and qualitative modifications to the contents of consciousness, including their experienced intensity \cite{revonsuo2009altered}. Third, at least some states can be ordered along a hierarchy of levels, from states of "reduced" consciousness (e.g. general anaesthesia, sleep) to others considered "richer" (e.g. certain states achieved during meditation or induced by pharmacological means) \cite{carhart2014entropic}. 

A proper definition of what constitutes an altered state of consciousness is, unfortunately, more elusive than suggested by the examination of these examples. If states of consciousness are transient, then what is their minimum accepted length? Do qualitative modifications of conscious content apply only to the sensory domain, or encompass other forms of subjective experience as well? Does a \emph{déjà-vu} (a brief episode of eerie familiarity with an unknown past event) qualify as an altered state of consciousness? What about an orgasm, or the state of pain caused by hitting one's finger with a hammer? Without doubt, these examples modify in one way or another the general contents of consciousness, but they are not commonly considered as altered states of consciousness. 

The intuitive notion of "levels" of consciousness is also problematic \cite{bayne2016there}. We are familiar with the fact that some states appear to be "more conscious" than others; for instance, ordinary wakefulness would have a higher conscious level than deep sleep or an absence seizure. But in what sense is deep sleep more or less conscious than an absence seizure? Following this logic, how should dreaming, the acute effects of psychedelic drugs, and the state achieved by expert meditators be ordered along a hypothetical uni-dimensional hierarchy of levels of consciousness? It seems that altered states of consciousness can only be subject to partial ordering, with comparisons between certain pairs of states being questionable or outright meaningless. 

These difficulties relate to two main problems. The first problem is granularity: how long is long enough to qualify as an altered state of consciousness? The second is compositeness: instead of a single level of intensity, multiple dimensions are likely required for an unambiguous characterisation; however, it is unclear how many dimensions are needed and how they should be determined \cite{bayne2016there, bayne2018dimensions}. A subsidiary issue related to the granularity problem is whether altered states of consciousness represent discrete states with sharply defined boundaries, or are more adequately understood as continuous transitions.

Several proposals have been put forward to circumvent these issues and define altered states of consciousness \cite{vaitl2005psychobiology, revonsuo2009altered, overgaard2010neural,tassi2001defining, ludwig1966altered, tart1976_, bayne2007conscious}. Here, we adopt perforce a more pragmatic stance: we are interested in altered states of consciousness lasting enough to be investigated by modern neuroimaging techniques (>10 min). At the same time, we strive to show that whole-brain models can be sufficiently rich to transcend the unidimensional characterisation of consciousness in terms of "levels". 

For the purposes of this article, we divide altered states of consciousness into the following (neither exhaustive nor mutually exclusive) categories: natural or endogenous (e.g. the states within the sleep cycle), induced by pharmacological means (e.g. general anaesthesia, the psychedelic state), induced by other means (e.g. meditation, hypnosis), caused by pathological processes, either neurological or psychiatric (e.g. disorders of consciousness, epilepsy, psychotic episodes), and transitory vs. permanent.

\begin{table}[t]
\centering
\caption{Categories of altered states of consciousness}
\label{tab:categories}
\def\arraystretch{2.0}
\setlength\tabcolsep{0.5cm}
\begin{tabular}{ c|c|c } 
\textbf{Category} & \textbf{Examples} & \textbf{Reversibility} \\ 
\toprule
Natural or endogenous & \makecell{deep sleep\\dreaming} & transitory \\ 
 Pharmacological & \makecell{general anaesthesia\\psychedelic state} & transitory\\
 Induced by other means & \makecell{meditation\\hypnosis} & transitory \\
 Pathological & \makecell{epilepsy\\psychotic episodes} & transitory or permanent \\ 
\end{tabular}
\end{table}

\section{Top-down signatures of consciousness from brain signals}
\label{sec:data-driven_tools}

A major challenge in the study of altered states of consciousness has been to establish empirical signatures in brain signals that are characteristic of different states, thus allowing us to identify them "from the outside" -- i.e. not depending on self-report or behavioural tasks \cite{sergent2012imaging}. Establishing and validating these signatures also carries significance from a clinical perspective, since they could lead to efficient and specific biomarkers for certain neuropsychiatric conditions \cite{sitt2014large, qutncy2004neural}. Furthermore, when interpreted within a broader theory, some of these signatures may also provide new insights about the nature of the corresponding conscious states, advancing our fundamental understanding of consciousness itself.

In the following, we first provide a broad overview of general aspects of theories of consciousness, and then illustrate what a signature of consciousness is by reviewing two well-known examples.

\subsection{Functionalist and non-functionalist positions on the mind-brain problem}

When we consider the most prominent contemporary theories of consciousness, we find that they mainly differ in what they take as valid empirical data to be explained by the theory. There are essentially two positions on this matter, which can be related to the influential division between functionalist and non-functionalist positions on the mind-brain problem. For a functionalist, the subjective quality of conscious experience is rejected as a valid target of scientific explanation. According to this view, most famously articulated by Daniel Dennett in \emph{Consciousness Explained}~\cite{Dennet1997}, only third-person objective measurements fall into the scope of a science of consciousness. This data is limited to observable behaviour and neural activity recordings; for instance, whenever an experimental subject claims to be experiencing a certain shade of blue, the neuroscientist is not tasked with finding how a physical process in the brain can cause a subjective feeling of blue, but with determining the mechanisms leading the subject to declare such experience \cite{dennett2003s}. Non-functionalists, on the other hand, reject this position as a sophisticated form of behaviourism \cite{block1978troubles}. According to this view, introspection plays a crucial role in the scientific explanation of consciousness, because it reveals the very nature of the \emph{explanandum} itself; any other kind of data represents, at best, an indirect approximation \cite{lutz2002guiding, shear1999view, chalmers1999first}. It is one of the defining features of consciousness, argue the defenders of this position, that it cannot be illusory \cite{frankish2016illusionism} since being conscious of something is precisely what bears that conscious experience into existence \cite{nida2016illusion, seager2017could}. 

When translated into the domain of neuroscience, these positions inform the two most influential contemporary models of consciousness. The global neuronal workspace theory (GNW) \cite{Baars2005, Dehaene2001} links consciousness with the widespread and sustained propagation of activity in the cortex, serving the computational function of broadcasting information to be processed by specialised modules \cite{mashour2020conscious}. This theory was developed to explain the neural signatures of consciousness seen in cognitive neuroscience experiments -- in other words, to explain third-person objective data. On the contrary, integrated information theory (IIT) \cite{Tononi2004, Balduzzi2008, Oizumi2014} is based on certain first-person qualities of subjective experience, which are accessed by introspection and can be taken as "postulates" or "axioms" for the theory \cite{Oizumi2014}. This theory strives to provide a quantitative characterisation of consciousness, as well as to determine the neural correlates of conscious contents from first principles only (even though concrete predictions may be computationally intractable~\cite{Barrett2019}). Both theories have been the target of intense criticism \cite{block2011perceptual, tsuchiya2015no, aru2012distilling, lamme2006towards, doerig2019unfolding, tsuchiya2019reply}, which can be taken as a sign that the scientific problem of consciousness remains unsolved.

While GWT and IIT are frequently pitted against each other, their predictions for human brains may still be mutually compatible ~\cite{seth2006theories, tagliazucchi2017signatures}. For our purpose, what these two theories have in common is that they follow a top-down approach, in the sense that they both focus on abstract computational or information-theoretical principles, without necessarily specifying how these principles arise as a consequence of local dynamics within the underlying neural substrate. We argue that it is via detailed whole-brain modelling that the points of agreement and divergence between theories, and how they relate to the neurophysiology of the human brain, can and should be studied ahead of possible experiments.

\subsection{Examples of signatures of consciousness} 

Since the conception of NCC, neuroscientists have turned to every available neuroimaging technology in the search for signatures of consciousness~\cite{Crick1990, Crick2003}. Although many kinds of signatures have been explored (including some related to metabolic consumption~\cite{Laureys2004} or cortical connectivity~\cite{Laureys1999}), for the purposes of this article we will focus on signatures measurable with functional neuroimaging tools like MEG, EEG and fMRI (which can be simulated with the models described in Section~\ref{sec:wholebrainmodels}). In the sequel, we illustrate the nature and application of signatures of consciousness by elaborating on two well-known examples.

\subsubsection{The entropic brain hypothesis}

One of the simplest, yet remarkably powerful, theoretical framework to furnish signatures of consciousness is Carhart-Harris' entropic brain hypothesis (EBH)~\cite{carhart2014entropic,carhart2018entropic}. According to the EBH,
the richness of conscious experience depends on the complexity of the underlying population-level neuronal activity, which determines the repertoire of states available for the brain to explore. 
Put simply, 
conscious states that involve richer experiences 
might require a more diverse set of brain configurations, which should leave a traceable footprint to be observed in the entropy, or in the entropy rate\footnote{While the entropy estimates the average uncertainty in a signal, the entropy rate estimates how hard is to predict the next time-point given its history.} of the corresponding brain signals.
Following this rationale, the level of consciousness should be proportional (at least within reasonable range) to the entropy of brain signals.

An effective tool to estimate the entropy rate of a signal is the Lempel-Ziv complexity (LZc) \cite{Lempel1976,ziv1978coding,carhart2018entropic}, originally conceived as a lossless compression algorithm.
The LZc of brain signals has proven to be an extremely robust signature of consciousness, and has been tested in a breadth of scenarios including
anaesthesia~\citep{zhang2001eeg}, coma~\cite{Nenadovic2014}, sleep~\citep{schartner2017global}, epilepsy~\cite{Dominguez2005}, meditation \cite{Vivot2020} and the psychedelic state~\citep{schartner2017increased,timmermann2019neural}. More recently, it has also been used to assess fluctuations of consciousness during normal wakefulness due to cognitive tasks~\cite{stam2005nonlinear}, stress~\cite{peng2013method}, fatigue~\cite{xu2018physical}, and music performance or listening~\citep{dolan2018improvisational}.

With its impressive track record and wide applicability, LZc stands as a prominent signature of consciousness to compare across biological and simulated brains. Furthermore, LZc can be used in tandem with transcranial magnetic stimulation to compute the perturbational complexity index~\cite{Casali2013}, a clinically-tested marker of consciousness, which can also be used as a test measure for whole-brain models.

\subsubsection{Integrated information theory}

A strong limitation of standard brain entropy analyses is that they consider only the entropy of individual signals, without acknowledging the multivariate structure of brain dynamics. An attractive way of studying interdependencies between brain signals is with tools drawn from integrated information theory (IIT)~\cite{Tononi2008}. IIT proposes an intimate relationship between consciousness and the ability of a physical system to 
be integrated in such a way that is "more that the sum of its parts" -- i.e. to display dynamical properties in the whole that are not observed in any of its parts. 

IIT builds on key information-theoretic ideas first presented in the seminal early work of Tononi, Sporns, and Edelman~\cite{Tononi1994}, and has been subject of continuous development since~\cite{Tononi2004, Balduzzi2008, Oizumi2014, mediano2019beyond}. Following Mediano \emph{et al.}~\cite{Mediano2019}, we distinguish between \emph{empirical IIT} and \emph{fundamentalist IIT} as two separate branches of the theory. While fundamentalist IIT has been highly controversial and subject of extensive criticism~\cite{mindt2017problem,morch2019consciousness,Barrett2019, bayne2018axiomatic}, multiple efforts in empirical IIT have been made to overcome the computational challenges of the theory~\cite{Krohn2017,kitazono2018efficient,toker2019information}.

At the core of empirical applications of IIT is a quantitative measure of integrated information, typically denoted by $\Phi$. There is currently no agreed-upon $\Phi$ measure, although multiple proposals have been put forward~\cite{Mediano2019} and can be used to understand and compare the dynamical structure of systems of interest. Detailed procedures describing how to compute different versions of $\Phi$ can be found in Ref.~\cite{Mediano2019}. Although the evidence supporting IIT as a fundamental theory of consciousness has been contested~\cite{Mediano2020}, measures inspired by empirical IIT have proven useful in analysing both empirical~\cite{Chang2012,Kim2018} as well as simulated~\cite{toker2019information,mediano2016integrated} neural data. Altogether, the family of information-theoretic measures inspired by empirical IIT provides a valuable toolkit to study the multivariate dynamics of whole-brain models.

\section{Bottom-up whole-brain models}\label{sec:wholebrainmodels}

While human neuroscience research has been increasingly dominated by imaging experiments, an important complement to this research is provided by computational neuroscience \cite{gerstner2012theory}. In effect, neuroimaging data is usually insufficient to inform the underlying mechanisms at play behind neural phenomena unfolding at different spatial and temporal scales \cite{ramsey2010six}. Also, since ethical considerations severely limit direct causal manipulation of human brain activity, most of the neuroimaging literature is limited to correlational studies.

The application of computational models to neuroimaging data with the purpose of making causal and mechanistic assertions has been proposed and developed in parallel with different objectives. For instance, deep neural networks can be used to model information-processing in the brain \cite{kriegeskorte2018cognitive} by comparing their representational content via second-order isomorphisms (e.g. representational similarity analysis) \cite{kriegeskorte2008representational}. These models can be used to investigate the plausibility of different computational architectures within cognitive neuroscience \cite{kriegeskorte2013representational}. Another example is dynamic causal modelling (DCM), which was developed to make model-based causal inferences from neuroimaging experiments \cite{friston2003dynamic}. DCM is based on simulating brain signals under the assumption of different causal interactions and then performing model comparison and selection. Finally, whole-brain models are based on dynamical systems coupled by large-scale anatomical connectivity networks, and are developed to reproduce the statistics of empirical brain signals at multiple scales \cite{schirner2018inferring}. We also distinguish whole-brain models from attempts to produce extremely detailed reproductions of large neural circuits (e.g. cortical columns) \cite{markram2006blue}, mainly due to differences in model complexity.

Whole-brain models provide a practical, ethical, and inexpensive "digital scalpel", which allows researchers to explore the counterfactual consequences of modifying structural or dynamical aspects of 
the brain. More generally, whole-brain models build a bridge from local networked dynamics to the large-scale patterns of activity that are addressed by theoretical signatures of consciousness. As such, they represent a valuable tool to narrow the space of mechanistic explanations compatible with the observed neuroimaging data, including data acquired from subjects undergoing different altered states of consciousness.

In this section, we provide a brief introduction to whole-brain models to the unfamiliar reader, discussing their various types and the principles behind their tuning to empirical data. Additionally, we review recent articles where these models have been used to shed light on the neurobiological mechanisms underlying different altered states of consciousness.

\begin{figure}[t]
\includegraphics[width=\linewidth]{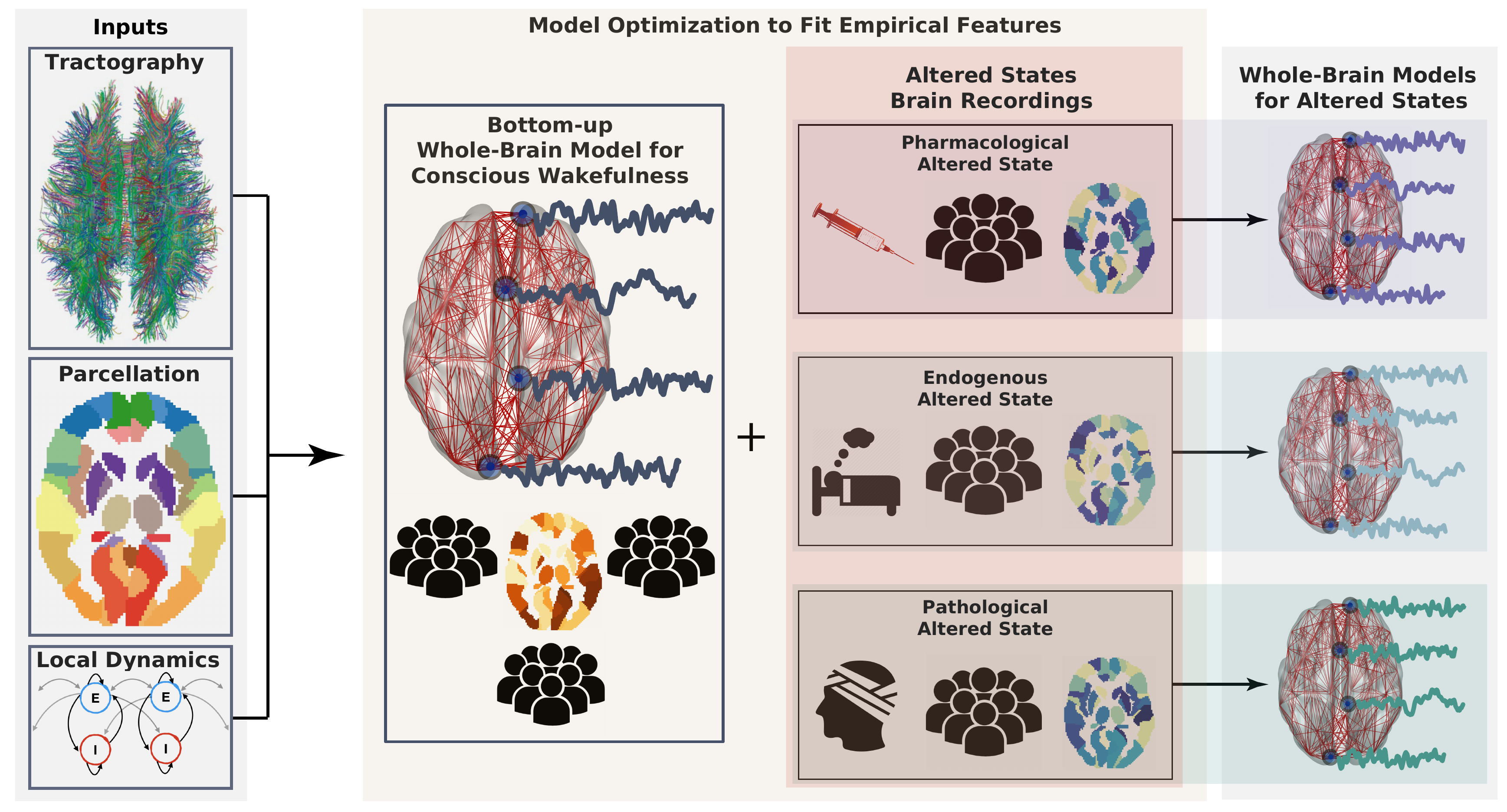}
\label{bottom-up}
\caption{
Workflow describing the construction of whole-brain models. First, model inputs are determined based on anatomical connectivity, a brain parcellation (representing a certain coarse graining), and the local dynamics (left).
Each region defined by the parcellation is endowed with a specific connectivity profile and local dynamics.
Then, the model can be optimised to generate data as similar as possible to the brain activity observed during conscious wakefulness. Generally, this similarity is determined by certain statistical properties of the empirical brain signals, which constitute the target observable. 
The same or another observable is obtained from subjects during altered states of consciousness and used again as the target of an optimisation algorithm to infer model parameters.
Following a given working hypothesis, the model for wakeful consciousness can be perturbed in such a way that optimises the similarity between the target observable for the altered state of consciousness and the data generated by the model.
In this way, a whole-brain model for an altered state of consciousness can be used to test working hypotheses about its mechanistic underpinnings. 
}
\end{figure}

\subsection{What are whole-brain models?}

Whole-brain models are sets of equations that describe the dynamics and interactions between neural populations in different brain regions. These models typically focus on the joint evolution of a set of key biophysical variables using systems of coupled differential equations (although discrete time step models can also be used, as will be discussed below). These equations can be built from knowledge concerning the biophysical mechanisms underlying different forms of brain activity, or as phenomenological models chosen by the kind of dynamics they produce. Then, local dynamics are combined by \emph{in vivo} estimates of anatomical connectivity networks. In particular, fMRI, EEG and MEG signals can be used to define the statistical observables, diffusion tensor imaging (DTI) can provide information about the structural connectivity between brain regions by means of whole-brain tractography, and PET imaging can inform on metabolism and produce receptor density maps for a given neuromodulator.

Most whole-brain models are structured around three basic elements:

\begin{enumerate}
   
\item[\textbf{A.}] \textbf{Brain parcellation:} A brain parcellation determines the number of regions and the spatial resolution at which the brain dynamics take place. The parcellation may include cortical, sub-cortical, and cerebellar regions. Examples of well-known parcellations are the Hagmann parcellation ~\cite{Hagmann2008}, and the automated anatomical labeling (AAL) atlas \cite{Tzourio-Mazoyer2002}.

\item[\textbf{B.}] \textbf{Anatomical connectivity matrix:} This matrix defines the network of connections between brain regions. Most studies are based on the human connectome, obtained by estimating the number of white-matter fibers connecting brain areas from DTI data combined with probabilistic tractography \cite{sporns2005human}. For control purposes, randomized versions of the connectome (null hypothesis networks) may also be employed.

\item[\textbf{C.}] \textbf{Local dynamics:} The activity of each brain region is typically determined by the chosen local dynamics plus interaction terms with other regions. A variety of approaches have been proposed to model whole-brain dynamics, including cellular automata \cite{Tagliazucchi2016, Haimovici2013}, the Ising spin model \cite{Deco2012, marinazzo2014information, abeyasinghe2020consciousness}, autoregressive models \cite{Messe2014}, stochastic linear models \cite{Saggio2016}, non-linear oscillators \cite{Cabral2014, Jobst2017}, neural field theory \cite{Robinson2015, BabaieJanvier2018}, neural mass models \cite{Breakspear2003, Honey2009}, and dynamic mean field models \cite{Deco2014, Deco2018, Kringelbach201921475}. A detailed review of the different models that can be explored within this context can be found in \cite{breakspear2017dynamic, deco2008dynamic}.

\end{enumerate}

The first two items are guided by available experimental data. In contrast, the choice of local dynamics is usually driven by the phenomena under study and the epistemological context at which the modelling effort takes place. Because of this hybrid nature, whole-brain models constructed following this process are sometimes called \emph{semi-empirical} models. Whole-brain models can be constructed from in-house code, or more easily from platforms such as The Virtual Brain (\url{https://www.thevirtualbrain.org/tvb/zwei}) \cite{ritter2013virtual}.

\subsection{Examples} 

We showcase two models that have been frequently used to assess mechanistic hypotheses behind both pharmacologically and physiologically-induced altered states of consciousness: the dynamic mean field model~\cite{Deco2014, Deco2018, Deco2019}, and the model comprised by Stuart-Landau non-linear coupled oscillators ~\cite{Cabral2014, Jobst2017, Ipina2020}. These examples are chosen to represent a biologically realistic model (dynamic mean field) and a phenomenological model (Stuart-Landau oscillators); moreover, these models have been applied to different states of consciousness, making them pertinent in the context of the present discussion.

\subsubsection{Dynamic mean field (DMF) model} 

In this approach, the neuronal activity in a given brain region is represented by a set of differential equations describing the interaction between inhibitory and excitatory pools of neurons  \cite{renart2004mean}.
The DMF presents three variables for each population: the synaptic current, the firing rate, and the synaptic gating, where the excitatory coupling is mediated by NMDA receptors and the inhibitory by GABA-A receptors. The interregional coupling is considered excitatory-to-excitatory only, and a feedback inhibition control in the excitatory current equation is included~\cite{Deco2014}. The output variable of the model is the firing rate of the excitatory population that is then included in a nonlinear hemodynamical model \cite{friston2000nonlinear} to simulate the regional BOLD signals.

The key idea behind the mean-field approximation is to reduce the high-dimensional randomly interacting elements to a system of elements treated as independent.
Then, an average external field
effectively replaces the interaction with all other elements. 
Thus, this approach represents the average activity of an homogeneous population of neurons by the activity of a single unit of this class, reducing in this way the dimensionality of the system. In spite of these approximations, the dynamic mean field model incorporates a detailed biophysical description of the local dynamics, which increases the interpretability of the model parameters.

\subsubsection{Stuart-Landau non-linear oscillator model}

This approach builds on the idea that
neural activity can exhibit -- under suitable conditions -- self-sustained oscillations at the population level~\cite{Cabral2014,Jobst2017,Ipina2020,Tagliazucchi2016, Deco2018b}.
In this model, the dynamical behaviour is represented by a non-linear oscillator with the addition of Gaussian noise at the proximity of a Hopf bifurcation \cite{marsden2012hopf}. By changing a single model parameter (i.e. bifurcation parameter) across a critical value, the model gives rise to three qualitatively different asymptotic behaviours: harmonic oscillations, fixed point dynamics governed by noise, and intermittent complex oscillations when the bifurcation parameter is close to the bifurcation (i.e. at dynamical criticality). Correspondingly, the model is determined by two parameters: the bifurcation parameter of the Hopf bifurcation in the local dynamics, and the coupling strength factor that scales the anatomical connectivity matrix. In contrast to the DMF model, coupled Stuart-Landau non-linear oscillators constitute a phenomenological model, i.e. the model parameter does not map into any biophysically relevant variable. In this case, the model is attractive due to its conceptual simplicity, which is given by its capacity to produce three qualitatively different behaviours of interest by changing a single parameter.

\subsection{How to fit whole-brain models to neuroimaging data? } 

Whole-brain models are tuned to reproduce specific features of brain activity. The way in which this is ensured is via optimisation of the free parameters in the local dynamics plus the coupling strength. Parameter values are usually selected such that the model matches a certain statistical observable computed from the experimental data. 

For example, the DMF whole-brain model introduces one parameter to scale the strength of the connectivity matrix, usually known as the \emph{global coupling parameter}. During model training, an exhaustive exploration of this parameter is conducted over a wide range of values. The parameter value is chosen to maximise the similarity between the observable computed from simulated and experimental data. For instance, the parameter can be chosen to minimise the Kolmogorov-Smirnov distance between the functional connectivity dynamics (FCD) distributions of the simulated and real data \cite{Deco2014}.

This kind of brute-force optimisation is employed when the number of free parameters is low (i.e. two or three). However, it is also possible to separately optimise the parameters governing the local dynamics of each node, which dramatically increases the dimensionality of the search space, and thus requires more elaborated optimisation techniques, such as gradient descent~\cite{Deco2017} or genetic algorithms~\cite{Ipina2020}. 
The advantage of considering a small set of global parameters resides in its simplicity and scalability, but unfortunately it misses the dynamical heterogeneity among brain regions. These heterogeneities can be modelled at the expense of increasing the parameter space. Essentially, the choice of model complexity (i.e. the number of free parameters) depends on the scientific question and its associated hypotheses.

Since adding more free parameters increases the computational cost of the optimisation procedure, it becomes critical to choose parameters reflecting variables that are considered relevant, either from a general neurobiological perspective or in the specific context of the altered state under investigation.
Depending on the latter, the parameters could be divided into groups that are allowed to change independently based on different criteria, including structural lesion maps, receptor densities, local gene expression profiles, and parcellations that reflect the neural substrate of certain cognitive functions, among others.

After choosing the parcellation, the equations governing the local dynamics and their interaction terms, the interregional coupling given by the structural connectivity matrix, and selecting a criteria to constrain the dimensionality of the parameter space, the last critical step is to define the observable which will be used to construct the target function for the optimisation procedure. As mentioned above, one possibility is to optimise the model to reproduce the statistics of functional connectivity dynamics (FCD). Perhaps a more straightforward option is to optimise the "static" functional connectivity matrix computed over the duration of the complete experiment, an approach followed by Refs.~\cite{Ipina2020} and ~\cite{Jobst2017}, among others. Other observables related to the collective dynamics can be obtained from the synchrony and metastability, as defined in the context of the Kuramoto model \cite{Jobst2017,shanahan2010metastable}. In general, any meaningful computation summarising the spatiotemporal structure of a neuroimaging dataset constitutes a valid observable, with the adequate choice depending on the scientific question and the altered state of consciousness under study. 

Since different observables can be defined, reflecting both stationary and dynamic aspects of brain activity, a natural question arises: is a given whole-brain model capable of simultaneously reproducing multiple observables within reasonable accuracy? We consider this question to be very relevant, yet at the same time it has been comparatively understudied. For instance, a review of articles using coupled Stuart-Landau oscillators shows that dynamical observables are reproduced when the oscillators operate at dynamical criticality (i.e. near the Hopf bifurcation), yet stationary observables (such as the "static" functional connectivity") are best reproduced for other parameter combinations \cite{Ipina2020,  Jobst2017, Deco2017}. This suggests that exploring bifurcations with higher co-dimensions or even chaotic dynamics unfolding in the proximity of strange attractors could enable the simultaneous optimisation of several observables, a possibility that is discussed later in this article.

Finally, some natural candidates for observables to be fitted by whole-brain models are precisely the high-level signatures of consciousness put forward by theoretical predictions, such as the different measures of information integration, complexity and entropy that were reviewed in the previous section. The objective is to fit whole-brain models using these signatures as target functions and then assess the biological plausibility of the optimal model parameters, which allows to test the consistency of these signatures from a bottom-up perspective. Alternatively, signatures of consciousness can be computed from the model --initially fitted to other observables-- and compared to the empirical results. Again, this highlights the need to understand which kind of local dynamics allow the simultaneous reproduction of multiple observables derived from experimental data.

\subsection{Whole-brain models applied to the study of consciousness}

The available evidence suggests that 
states of consciousness are not determined by activity in individual brain areas, but emerge as a global property of the brain, which in turn is shaped by its large-scale structural and functional organisation~\cite{Dehaene2001, Dehaene2011, tononi1998consciousness}. According to this view, whole-brain models provide a fertile ground to explore how global signatures of different states of consciousness emerge from local dynamics. This promise is already being met, as shown by several recent articles \cite{Cabral2014,Jobst2017,Ipina2020,Tagliazucchi2016, Deco2018b, Deco2019, carhart2014entropic, Bocaccio2019}. 

For example, transitions from wakefulness into other states, such as the different stages of human sleep or the state induced by general anaesthetics, have been interpreted as phase transitions in neural mass models and in terms of the collective dynamics of coupled Stuart-Landau oscillators \cite{Cabral2014,Jobst2017,Ipina2020}. Noise-driven systems at dynamical criticality result in dynamics compatible with neuroimaging recordings obtained during conscious wakefulness, and departures from these dynamics better reflect different states of unconsciousness \cite{Tagliazucchi2016, Deco2018b, carhart2014entropic, Bocaccio2019, solovey2015loss, alonso2014dynamical}. As will be discussed in the following section, the stochastic switching between different attractors results in the kind of metastable behaviour that is characteristic of conscious wakefulness \cite{cavanna2018dynamic}. These results are consistent with the hypothesis of statistical criticality (e.g. proximity to a second order phase transition) as a fundamental principle of brain organization \cite{Chialvo2010}. Even though parallels can be drawn between statistical and dynamical criticality, we limit our discussion to the former since the relationship between both concepts is complicated and beyond the scope of this article.

Following the example of the PCI index (which is obtained by perturbing the cortex with TMS and measuring the complexity of the elicited response) \cite{Casali2013}, whole-brain models can be systematically "perturbed" by incorporating changes into the dynamical equations. The \emph{in silico} rehearsal of perturbations is useful to test hypotheses concerning which parts of the model are essential to produce different signatures of consciousness. A prominent example of this perturbational analysis applied to whole-brain models can be found in a recent article~\cite{Deco2019} where a whole-brain model based on coupled Stuart-Landau oscillators was fitted to empirical fMRI data acquired from subjects during deep sleep. The model was then modified by changing local bifurcation parameters with a greedy optimization algorithm, which unveiled the optimal perturbation profile to increase the similarity to a target brain state (in this case, conscious wakefulness). Another relevant example of this perturbational approach is found in Ref.~\cite{Deco2018}, where a transition was shaped by the effects of neuromodulation. The authors investigated the transition from resting state activity acquired under a placebo condition towards the altered state of consciousness induced by the serotonin 2A receptor agonist lysergic acid diethylamide (LSD). A dynamical mean-field model was fitted to minimize the difference between FCD of the simulated activity and the empirical data of subjects in the placebo condition, which allowed to determine the optimal value of the global coupling parameter. Then, an empirical map of 5-$HT_{2A}$ receptor density was used to modulate the synaptic gain, effectively simulating the heterogeneous effects of LSD across the whole brain. As a control, the authors showed that using maps for the density of other serotonin receptor sub-types decreased the goodness of fit, thus corroborating the well-known association between LSD and the 5-$HT_{2A}$ receptor.

Another interesting possibility is to assess the consequences of stimulation protocols that are impossible to apply \emph{in vivo}. An example is the Perturbative Integration Latency Index (PILI) \cite{Deco2018b}, which measures the latency of the return to baseline after a strong perturbation that generates dynamical changes detectable over long temporal scales (on the order of tens of seconds). This \emph{in silico} perturbative approach allows to systematically investigate how the response of brain activity upon external perturbations is indicative of the state of consciousness, providing new mechanistic insights into the capacity of the human brain to integrate and segregate information over different time scales.

In Ref.~\cite{Ipina2020}, the authors used a model of coupled Stuart-Landau oscillators to model the regional changes in dynamical stability that occur during the wake-sleep cycle. Brain regions belonging to different resting state networks (RSN) \cite{damoiseaux2006consistent} were considered as independent sources of variation for the local model parameters. Using a stochastic optimisation algorithm, the authors represented the transition from wakefulness into deep sleep as a sequence of changes in the stability of brain activity within canonical RSN. A follow-up paper extended this analysis to other states of reduced consciousness (including anaesthesia and patients suffering from disorders of consciousness) and investigated the possibility of inducing transitions to conscious wakefulness by means of simulated periodic stimulation at the resonant frequency of each node in the model \cite{perl2020perturbations}.

\section{Proposed research agenda}
\label{sec:proposed_agenda}

\subsection{Motivation}

Consciousness research is in need of mechanistic accounts to explain why brain signals recorded during different states of consciousness can be consistently characterised by the presence of certain global signatures. Our motivation is not the replacement of the explanations of these signatures provided by theories such as GNW or IIT. Instead, we aim to put forward a framework for their investigation from a bottom-up perspective. Eventually, we expect to converge on the high-level explanations furnished by some of these theories.  Our inspiration is partially drawn from statistical thermodynamics, which provides a clear example of how the bottom-up and top-down perspectives can converge into a consistent picture of physical reality. Importantly, in this case the resulting theory remained useful both as a set of phenomenological principles and computational rules (i.e. classical thermodynamics), but also as a framework to establish connections between those principles and the rules governing the microscopic properties of matter.

Following this concept, we strive to use our current knowledge about neural dynamics to produce models whose behaviour agrees with the constraints of some  theories formulated from a top-down perspective, while weakening the support for others as a result of inconsistent predictions. Here it becomes important to clarify our intended meaning of the word "prediction". When it comes to complex systems such as the brain, predictions are considered possible only in a statistical sense \cite{Chialvo2010}. Accordingly, we do not expect that the time series generated by computational models directly correspond to their empirical counterparts; however, we can expect a match for statistical observables.

This motivates our study of altered states of consciousness, since their extended temporal duration guarantees the possibility of extracting robust statistical characterisations from multivariate neuroimaging recordings. An example of this characterisation is the matrix derived from computing all pairwise correlations between regional time series, which is considered a marker of inter-areal functional connectivity (sometimes referred to as the "functional connectome")~\cite{smith2013functional}. We consider that whole-brain computational models have been developed to a point where they contain sufficient empirical ingredients to predict the second-order statistics of brain activity. Thus, the field is ripe to welcome a framework which may provide solid ground to investigate signatures of consciousness from a mechanistic perspective. 

The following example is aimed to motivate the proposal we put forward in the next section. We know that activity within a network of brain regions including the fronto-parietal cortex is correlated with conscious experience~\cite{Laureys1999,Cavanna2006, Andersen2016, Koch2016, Utevsky2014}. On the other hand, conscious experience is also characterised by signatures such as information integration, entropy and neural complexity. Is it possible to determine the causal role that these anatomical regions play in the generation of these signatures of consciousness by means of computational models? 

\subsection{Proposal}

The principal idea behind our proposal is that whole-brain models can be used to test hypotheses concerning the mechanistic and causal underpinnings of different states of consciousness. We do not expect that whole-brain models are sufficiently advanced to identify those precise mechanisms; however, we propose that they can contribute to narrow the space of possible mechanistic explanations, therefore complementing current theories of consciousness from a bottom-up perspective.

The fundamental objective of this research program is to foster the development of this novel approach to study altered states of consciousness. Our framework rests upon the complementary nature of three key ingredients: experimental data obtained through neuroimaging experiments, theoretical approaches to characterise signatures of consciousness, and bottom-up whole-brain computational models. The application of modern neuroimaging techniques to the study of signatures of consciousness has provided very effective tools to predict the brain activity patterns that are associated with different states of consciousness. However, as Ren\'{e} Thom famously stated \emph{"to predict is not to explain"} \cite{thom1997}. Hence, we now turn to the discussion of how models could bridge the gap between prediction and explanation.

The proposed framework to model altered states of consciousness is based on adjusting three independent variables (see Figure~\ref{fig:triangle}):

\begin{figure}[t]
\includegraphics[width=\linewidth]{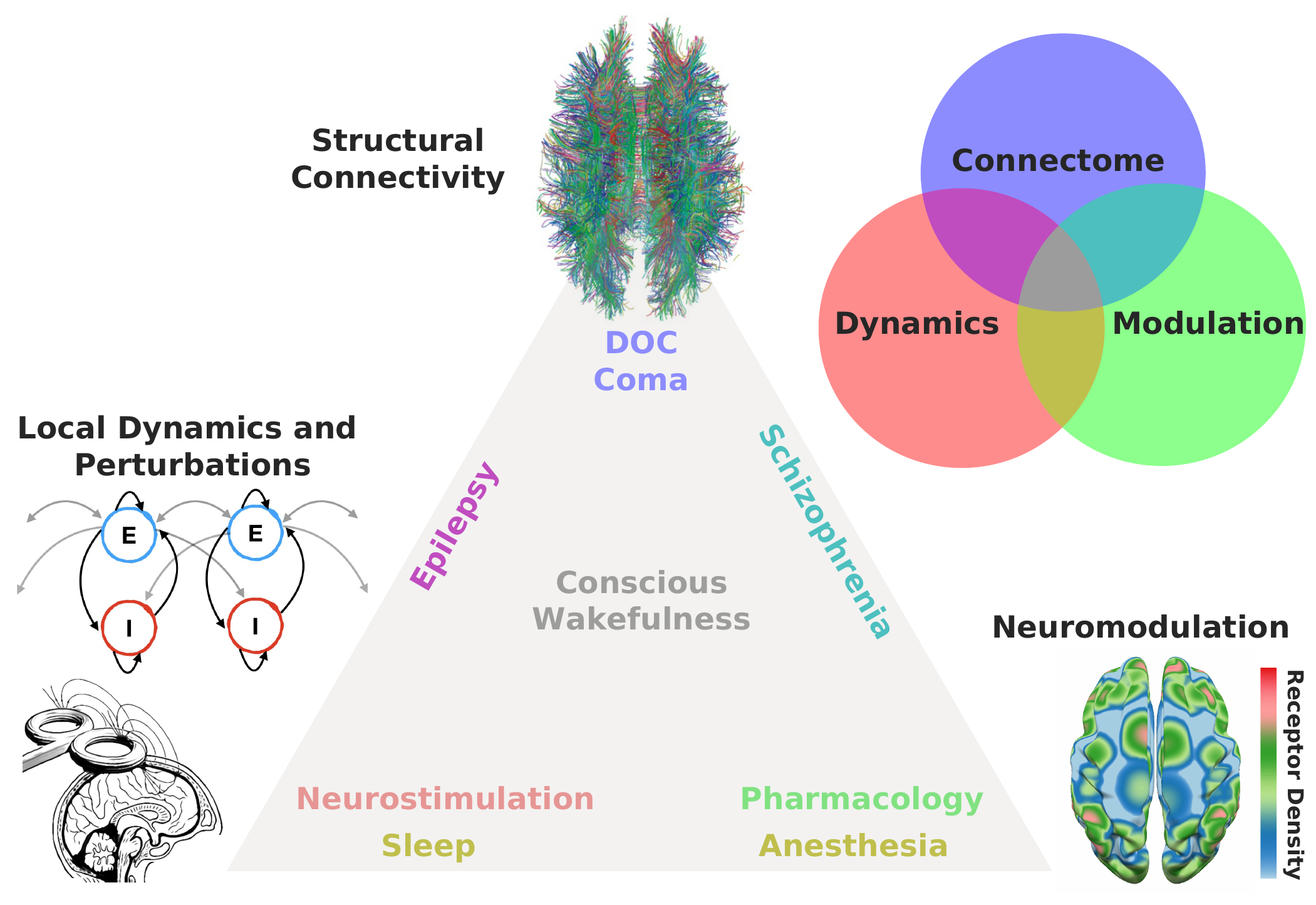}
\caption{\label{fig:triangle} Representation of the three key variables that can be modified to construct whole-brain models of different altered states of consciousness. These variables correspond to local dynamics, anatomical connectivity, and priors related to neuromodulatory systems necessary to accommodate physiological, pathological and pharmacologically-induced altered states of consciousness. Certain states may require the modification of multiple variables; for instance, focal seizures and propofol-induced anaesthesia are both associated with low complexity patterns of brain activity, yet in the first case these dynamics reflect structural abnormalities, while in the second case they reflect the activation of certain inhibitory pathways.}
\end{figure}

\begin{enumerate}
    \item[\textbf{A.}] \textbf{Connectome:} Is the state of consciousness implicated with local or diffuse structural abnormalities? This is frequently the case for neurological conditions such as coma and post-comatose disorders of consciousness (e.g. unresponsive wakefulness syndrome, minimally conscious state) \cite{fernandez2011diffusion}. Also, subtler structural modifications can be implicated in certain psychiatric conditions presenting episodes of altered consciousness, such as different forms of schizophrenia \cite{kubicki2005dti}.
    \item[\textbf{B.}] \textbf{Modulation:} Is the state of consciousness a consequence of neuromodulatory changes, either endogenous or induced externally by means of pharmacological manipulation? Two typical examples are the altered states of consciousness induced by psychedelics/dissociatives, which are linked to agonism/antagonism at serotonin/glutamate receptors \cite{nichols2016psychedelics}. Certain psychiatric conditions are believed to arise as a consequence of neuromodulatory imbalances, e.g. dopaminergic imbalances are believed to play an important role in the pathophysiology of schizophrenia \cite{howes2009dopamine}. Most anaesthetic drugs reduce the complexity of the brain activity by targeting specific neuromodulatory sites, such as those activated by gamma-aminobutyric acid (GABA) \cite{peduto1991biochemical}. Finally, sleep is a state of reduced consciousness triggered by activity in monoaminergic neurons with diffuse projections throughout the brain \cite{jouvet1972role}.
    \item[\textbf{C.}] \textbf{Dynamics:} Is the altered state of consciousness captured by well-understood dynamical mechanisms? Does the model include parametrically controlled external perturbations? While changes in the local excitation/inhibition balance are ultimately caused by neurochemical processes, they are best understood in terms of their dynamical consequences. States such as epilepsy, deep sleep and general anaesthesia are believed to involve unbalanced excitation/inhibition \cite{gao2017inferring}. In some cases, dynamics may be sufficiently idiosyncratic to be captured by low dimensional phenomenological models, as in the case of certain forms of epileptic activity \cite{el2020epileptor}. Finally, local dynamics could be modified to simulate the effects of external neurostimulation \cite{Deco2019, perl2020perturbations}.
    
\end{enumerate}

Depending on the answers to these questions, the whole-brain model should incorporate changes to anatomical connectivity, local dynamics, or include empirical receptor density maps to add a new layer of neurobiological detail. 

\subsection{What can we learn?}

The dynamics of whole-brain models can be perturbed arbitrarily. This is significant since it allows to explore different mechanisms leading to the observed empirical dynamics (as described in a previous paragraph) and to explore how external stimulation can force transitions between  states of consciousness, including the clinically relevant case of displacing whole-brain models from unconscious states towards wakefulness \cite{Deco2019, perl2020perturbations}. Therapeutic alternatives to accelerate the recovery of DOC patients are scarce, and while some studies support the therapeutic role of external electrical stimulation \cite{hermann2020combined}, very little is known about the optimal choice of stimulation sites and parameters. Whole-brain models could be useful for the optimization of stimulation protocols, as well as for assisting in clinical decision making. Localized stimulation and/or resection of neural tissue are surgical alternatives to treat certain severe forms of epilepsy, and whole-brain models have been explored with success to predict the outcome of these interventions \cite{an2019optimization}. The same concept could apply to the development and \emph{in silico} testing of new pharmaceuticals to treat psychiatric conditions, where whole-brain models could be used to reverse-engineer the optimal receptor affinity profiles required to restore statistical signatures of healthy brain dynamics. Finally, the combination of data produced by whole-brain models and machine learning classifiers could be useful for data augmentation in the context of automated diagnosis of rare neurological diseases \cite{perl2020data}, and to generate input for deep learning architectures (e.g. variational autoencoders) capable of representing altered states of consciousness as trajectories within a low dimensionality latent space. \cite{perl2020generative}.

\subsection{Case study: modelling neural entropy increases induced by psychedelics}

To further highlight what we can learn from whole-brain models, we discuss an illustrative example of a bottom-up model that successfully matches a global signature of altered conscious \cite{Herzog2020}. Using the DMF model optimised to fit the FCD of placebo and LSD conditions~\cite{Deco2018}, a significant entropy increase of brain signals was found in  LSD vs. placebo as a consequence of simulated 5-$HT_{2A}$ receptor activation. Thus, the model was capable of identifying a low-level (i.e. molecular scale) mechanism leading to increased neural entropy, which is a robust signature of the psychedelic state \cite{carhart2014entropic, carhart2018entropic}. 

Since activation of the 5-$HT_{2A}$ receptor is causally implicated with the conscious state induced by serotonergic psychedelics~\cite{nichols2016psychedelics,Kraehenmann2017,Preller2018}, the effect of the drug was modelled as a local change in the non-linearity of the regional firing rate. This change was proportional to the local density of 5-$HT_{2A}$ receptors as determined by PET imaging. Brain entropy increases during the psychedelic state were the result of heterogeneous changes in the entropy of the regional firing rates (i.e. some regions increased while others decreased their entropy). These changes in firing rate entropy depended both on the local anatomical connectivity and the 5-$HT_{2A}$ receptor density.

Thus, starting from local dynamics describing the behaviour of coupled excitatory and inhibitory pools of neurons, and introducing a perturbation which reflects serotonergic activation, the model provided a bottom-up confirmation of 5-$HT_{2A}$ activation as the source of increased neural entropy during the psychedelic state. In the context of Fig. \ref{fig:triangle}, the model adopted changes in local dynamics (bottom left) informed by empirical maps of 5-$HT_{2A}$ receptor density (bottom right).

\section{Future directions}

\subsection{What should be the "bottom" of bottom-up models?}
\label{sec:future}

The question of the ultimate substrate of consciousness is part of a long-standing philosophical debate, with positions including functionalism (the substrate is irrelevant insofar it instantiates the adequate set of causal relationships) \cite{Dennet1997}, biological naturalism (the view that consciousness arises as a consequence of biochemical processes in the brain) \cite{searle2007biological}, and proposals of consciousness as a manifestation of quantum mechanics \cite{stuart1998quantum}. Even though we choose to sidestep this complicated discussion, our modest aim of building bottom-up models of brain activity still requires the specification of some physical or biological substrate, which in turn determines the level of realism displayed by the equations that govern local dynamics.

Many signatures of consciousness are directly related to the global complexity of brain dynamics, reflecting the widespread hypothesis that consciousness plays an integrative role in the brain \cite{tononi1998consciousness}. According to this hypothesis, consciousness could be considered a dynamical process "gluing" together the output of specialised neural circuits. While tampering with these circuits could modify some specific contents of consciousness, only the disruption of large-scale neural communication would result in a state of altered or reduced consciousness. Since this view disregards the contribution of specific computations that are implemented in local neural circuitry, we could expect that bottom-up models capable of reproducing an adequate set of canonical dynamics\footnote{Here, \emph{canonical dynamics} refers to dynamics in the proximity of a class of topologically equivalent attractors. The reader should think of the result of simplifying the equations into the normal forms corresponding to the bifurcations present in the system \cite{murdock2006normal}.} will suffice to span the spectrum of signatures of altered consciousness. Conversely, it could be that the large-scale dynamics that support inter-areal communication at the same time interact and shape local information processing, and vice-versa. In this case, we expect that increasingly complex and biologically realistic models will be needed to advance with our proposal. 

This crucial point results in a ramification within our proposal to investigate altered states of consciousness using whole-brain models. On one hand, models could be enriched by increasingly detailed and sophisticated sources of empirical information with the purpose of linking signatures of consciousness to the biophysical details of neural activity. This direction is already suggested by studies modelling the effects of 5-$HT_{2A}$ activation using receptor density maps produced by PET imaging \cite{Herzog2020,Deco2018}. Following this direction, future models could be expanded to include fine-grained details of local wiring patterns, different cell types and their projections, as well as their interaction with diffuse neuromodulatory systems. However, as complexity is increased, the conceptual interpretation of models becomes less clear. On the other hand, it is known that dynamical systems may exhibit canonical behaviours when their solutions undergo changes in their qualitative behaviour (i.e. bifurcations) \cite{murdock2006normal}. Recent work fitting whole-brain models to the results of fMRI experiments suggests that bifurcations play a key role in the reproduction of the second-order statistics of empirical data \cite{Cabral2014,Jobst2017,Ipina2020,Tagliazucchi2016, Deco2018b}. This occurs because noisy dynamics close to a bifurcation point switches between different attractors, producing rich and complex dynamics typical of brain signals. This observation raises the question of whether more complex models reproduce the statistics of empirical observables by virtue of their universal behaviour near bifurcation points, or as a consequence of their stationary solutions away from dynamical criticality.

\subsection{Transitions between canonical dynamics as primitives to construct whole-brain models}
\label{sec:transitions}

Contrary to the \emph{dictum} by Norbert Wiener (\emph{"The best material model of a cat is another, or preferably the same, cat"}) we propose that even if vast sources of biological information can be incorporated into whole-brain models, striving for such level of detail defeats the purpose of unveiling concrete and interpretable mechanisms underlying signatures of consciousness. Thus, we suggest that models could be classified by the kind of large-scale activity patterns they are capable of generating. In other words, we propose that the "bottom" of bottom-up models should not be related to the scale of the biological substrate, but to the minimal set of simple dynamical behaviours necessary to reproduce a certain signature of consciousness. Paralleling the definition of NCC given by Crick and Koch ~\cite{Crick1990,Crick2003}, we could introduce the \emph{"dynamical correlates of consciousness"} (DCC); but we opt to not introduce yet another acronym in an already crowded field. 

Interestingly, Batterman has suggested that multiple realizability, the \emph{"metaphysical mystery"} that troubled Jerry Fodor, among other great philosophers of the mind, is as mysterious as the observation that physical matter behaves in ways which are entirely independent from the vast majority of its details~\cite{batterman2000multiple}. For a typical example consider a pendulum, whose behaviour is described by the same differential equation regardless of the colour of the swinging bob. Furthermore, in the small amplitude regime all systems with an U-shaped energy landscape can be approximated by an harmonic solution, with examples ranging from electrical circuits to orbital mechanics. Northoff and colleagues have argued that the spatiotemporal dynamics constitutes the fundamental substrate underlying human consciousness~\cite{northoff2019temporo}, which resonates with Batterman's proposal, as well as with our suggestion that the "bottom" (i.e. the maximum necessary level of detail) is best understood as a comprehensive list of the dynamical behaviours that the system can display. We postpone taking a stance towards these metaphysical speculations, and proceed to develop these ideas in the context of building useful bottom-up models in the future.

\begin{figure}[ht!]
\centering
\includegraphics[width=0.9\linewidth]{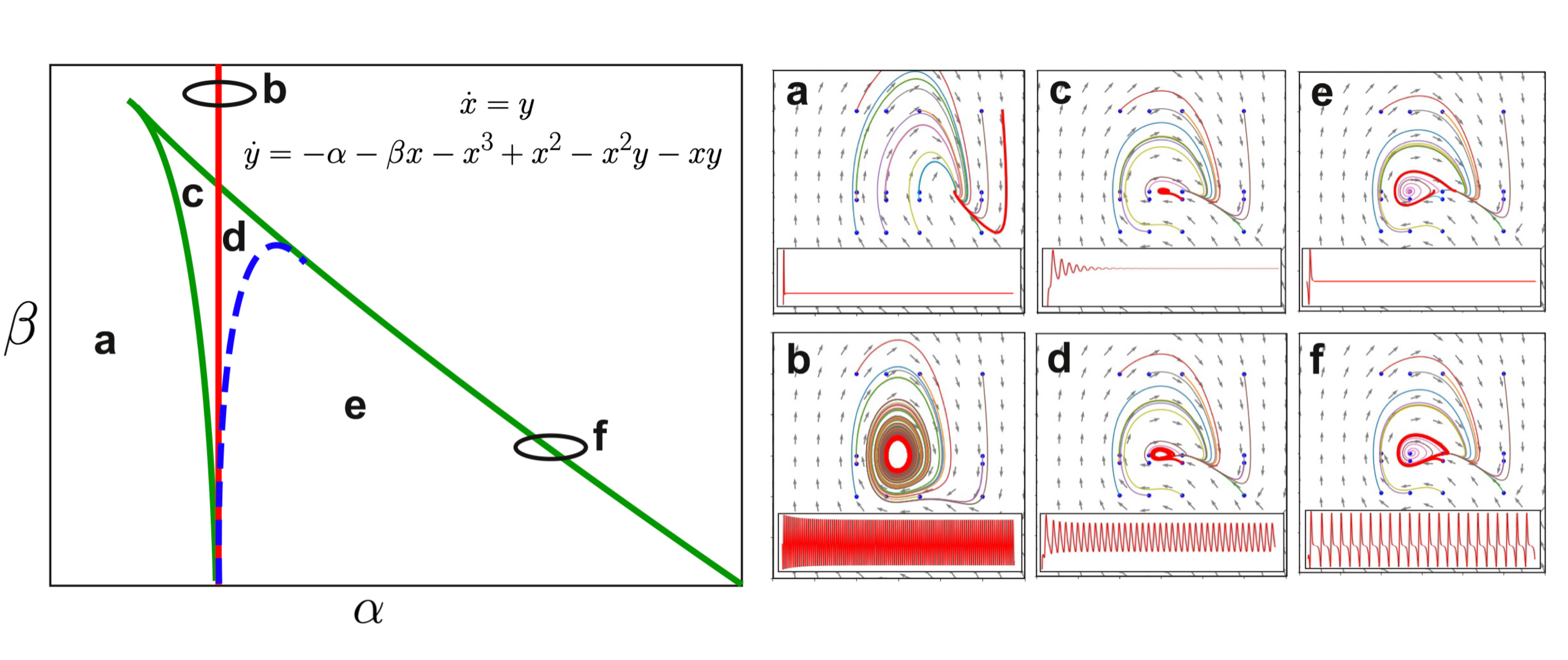}
\caption{\label{takens} \textit{Left panel:} Takens-Bogdanov bifurcation diagram, which is obtained by changing parameters $\alpha$ and $\beta$ in the normal form equations (included as an inset). Depending on the combination of parameters, this simple dynamical system can present qualitatively different solutions. The green line stands for a saddle-node bifurcation, where two equilibrium points collide and disappear. Crossing the red line results in a Hopf bifurcation, where dynamics switch from a fixed point to stable harmonic oscillations. The dashed line represents a homoclinic bifurcation, where the limit cycle collides with a saddle point resulting again in steady dynamics. \textit{Right panel:} The phase portraits \textbf{a-f} illustrate the dynamics at different regions of the bifurcation diagram, with individual trajectories highlighted in red and presented both as curves in phase space and as time series. a) Stable fixed point, b) Self-sustained harmonic oscillation after the appearance of a stable limit cycle, c) Three fixed points appear due to a saddle-node bifurcation, resulting in a stable fixed point, d) One of the stable fixed points loses its stability and dynamics undergo a Hopf bifurcation, e) The limit cycle undergoes a homoclinic bifurcation,  f) A saddle-node on a limit cycle (SNIC) bifurcation occurs, resulting in periodic dynamics with complex spectral content. For a detailed description of the Takens-Bogdanov bifurcation see Ref.~\cite{guckenheimer2007bogdanov}. Left panel adapted from Ref. \cite{gabo}.}
\end{figure}

A set of qualitatively different dynamics is provided in Fig. \ref{takens}, illustrating a Takens-Bogdanov bifurcation diagram \cite{guckenheimer2007bogdanov}. Whole-brain models can be constructed by coupling the dynamics given as an equation in the inset (left panel) either by variables $x$, $y$, or both. The equation and its solutions depend on two parameters, $\alpha$ and $\beta$. Under the weak coupling assumption, modifying these two parameters will result in qualitative changes in the local dynamics (where these changes occur in the diagram could be modified by the coupling strength). For uncoupled dynamics, parameter combinations at points  \textbf{a},  \textbf{c},  \textbf{e} result in a stable constant level of activity (i.e. fixed point dynamics). Parameter combinations at points \textbf{b},  \textbf{d},  \textbf{f} give rise to oscillations of different spectral content (i.e. limit cycles). 

In the right panel of Fig. \ref{takens}, the solutions can be visualised either as time series or as two dimensional diagrams known as phase portraits, where each axis corresponds to a variable (in this case, $x$ and $y$) and the arrows stand for the vector field (in this case, $\dot{x}$ and $\dot{y}$). Insofar the bifurcations in the left panel of Fig. \ref{takens} are not crossed, changes in the parameters  $\alpha$ and $\beta$ only result in deformations of the phase portrait, representing solutions that are equivalent in a qualitative sense (more formally, the phase portraits are topologically equivalent). Crossing a bifurcation results in an abrupt change that cannot be understood as a small deformation of the phase portrait, implying a qualitatively different behaviour of the system. 

The richness of coupling this kind of simple dynamical models stems from the possibility of inducing stochastic transitions across bifurcations by incorporating an additive noise term. In this way, dynamics switch intermittently between two qualitatively different solutions. In the case of the Hopf bifurcation, for instance, noise-driven dynamics at the bifurcation point are neither stable nor oscillatory, but present complex amplitude fluctuations \cite{Deco2017}. The noise-driven exploration of a system's attractor space is a mainstay of computational neuroscience \cite{rolls2010noisy} and could represent an useful methodological resource to build whole-brain models to explore altered states of consciousness. 

\begin{figure}[ht!]
\centering
\includegraphics[width=0.7\linewidth]{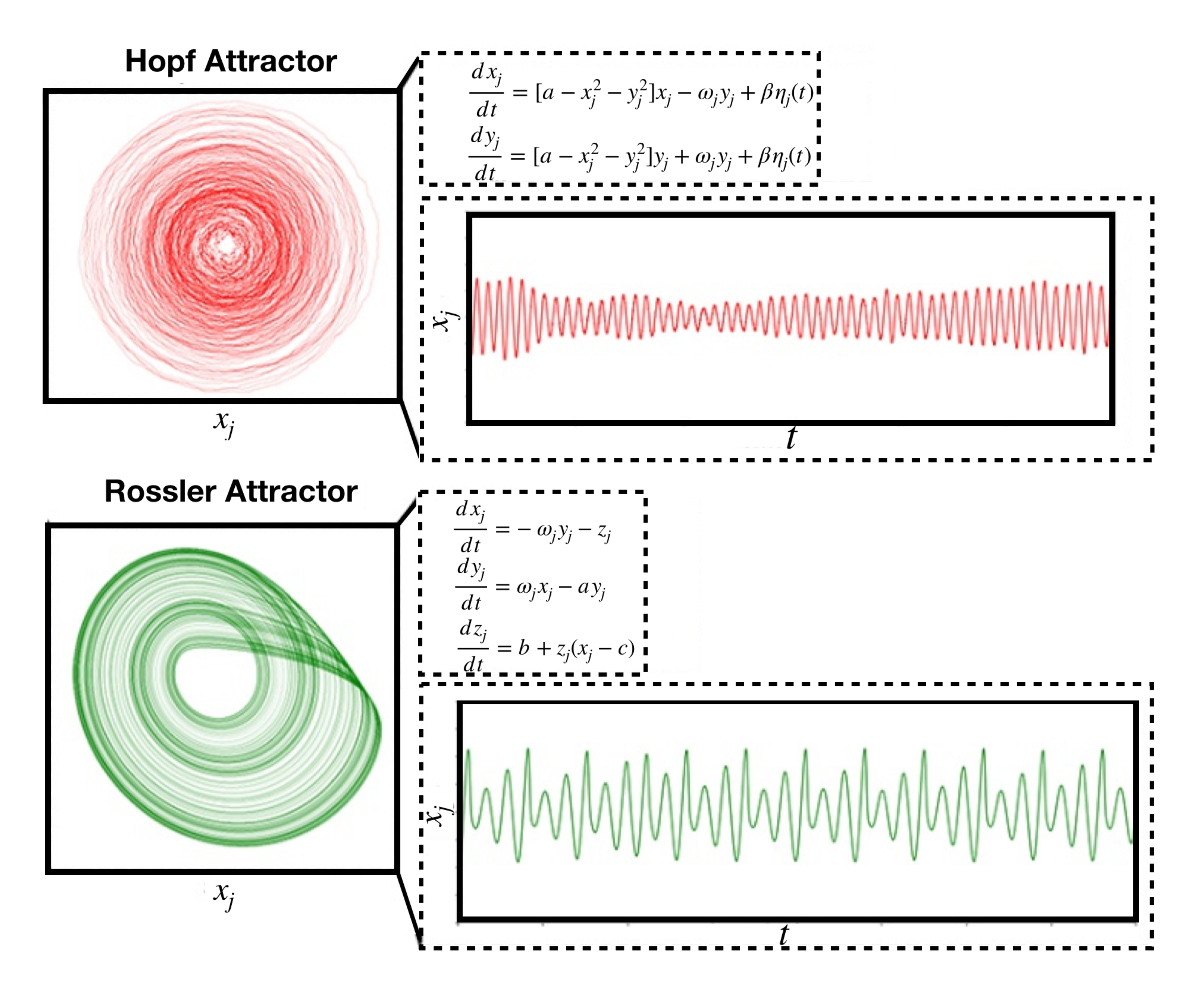}
\caption{\label{chaoshopf} \textit{Upper panel:} Phase space of a single Stuart-Landau nonlinear oscillator near dynamical criticality (Hopf bifurcation) with an additive noise term. The radius of the limit cycle fluctuates unpredictably, resulting in complex signal amplitude modulations. \textit{Bottom panel:} Phase space of a chaotic Rossler oscillation in a regime with positive Lyapunov exponent, without the addition of noise. Dynamics unfold in the proximity of a strange attractor, which results in complex but deterministic dynamics.}
\end{figure}

Following the pioneering work of Deco and colleagues \cite{Deco2017}, the most frequently explored transition is between stable noise-driven dynamics and self-sustained harmonic oscillations, corresponding to the Hopf bifurcation (vertical red line in Fig. \ref{takens}), which appears in Stuart-Landau nonlinear oscillators. At the bifurcation point, dynamics show the kind of complexity that is compatible with certain signatures of consciousness, with departures from this point being reported for states of reduced consciousness such as sleep and anaesthesia \cite{Deco2018b,Deco2019, Jobst2017, Ipina2020} (as it is clear from Fig. \ref{takens}, however, this bifurcation is only one among multiple possibilities). The upper panel of Fig. \ref{chaoshopf} illustrates this situation by presenting the phase space and temporal evolution of a noise-driven Stuart-Landau nonlinear oscillator near dynamical criticality. The signal evolves with complex amplitude fluctuations as noise drives the dynamics across the bifurcation. Also, at dynamical criticality small fluctuations tend to be amplified \cite{Deco2017, Jobst2017}, thus whole-brain models far from criticality reproduce the lack of sustained and complex responses to external perturbations seen in states of reduced consciousness \cite{Casali2013}. 

The inclusion of noise in whole-brain models raises questions concerning the mechanisms that endow biological systems with stochastic dynamics \cite{rolls2010noisy}. Again, we postpone these difficult questions \emph{in lieu} of more practical considerations, and propose that noise-driven equilibrium dynamics increase interpretability at the expense of two main shortcomings. First, parameter fine-tuning is required to pose dynamics near dynamical criticality. As discussed above, optimisation procedures can be applied to obtain the parameters which best reproduce certain empirical observables. However, the biological variables captured by the optimal combination of parameters could change upon small perturbations, leading to models that always predict intrinsically unstable states of consciousness. The second problem is that once parameters are optimised to reproduce a certain observable, other different observables could be poorly captured by the model, thus questioning the extent to which the model is adequately describing the empirical data. We propose that both problems could be simultaneously addressed by exploring non-stochastic models of chaotic coupled oscillators, such as Rossler oscillators. In this model, dynamics unfold near a strange attractor with positive Lyapunov exponent for a comparatively ample range of parameters \cite{letellier2006rossler}. Thus, complex dynamics do not depend on a bifurcation parameter taking a precise value, but instead arise over an extended range of parameter values. This kind of phenomenological models of whole-brain activity is comparatively understudied, and could represent a valuable target for future developments.

\section{Final remarks}
\label{sec:conclusion}

The history of science shows an intensive ongoing debate about the position of scientific inquires with respect to the study of consciousness. As a matter of fact, until recently the largest part of the scientific community did not consider consciousness as a suitable topic for investigation. While the ultimate nature of consciousness is still full of mysteries, it is evident that deepening our knowledge of the mechanistic, statistical, and dynamical relationships within the brain in its different possible states of consciousness can only increase our understanding of the relationship between mind and body.

A key factor supporting the modern discipline of consciousness research is the extraordinary development of neuroimaging technologies that occurred over the last decades, which plays a similar fundamental role than the one played by telescopes in the discovery of the nature of the solar system. However, making progress in the problem of consciousness not only depends on technological advances, but also on our capacity to explore and chart the contents and boundaries of consciousness itself. Consciousness research needs neuroimaging as much as any other branch of human neuroscience, but also needs to devise and explore new methods to induce altered states of consciousness, and to break through arbitrary regulatory restrictions preventing the exploration of certain older but very powerful research tools \cite{shulgin1992pihkal, shulgin1997tihkal}.  

These technological
advances, matched with increases in computational capability, and a renewed appreciation of the role that altered states of consciousness play in scientific research, have prepared a fertile ground for whole-brain models to open a new window of research possibilities.
In effect, while much progress has been made during the last decades in the problem of identifying top-down signatures of consciousness,
most of these tools have not yet reached a stage of maturity to allow clinical applications. We expect that pursuing the problem from a different perspective will be invigorating for the field as a whole, increasing the appreciation for the role that low-level biological mechanisms play in the emergence of high-level signatures of consciousness. 

Consciousness research is not alone in its need for low-level mechanistic explanations. The project of formulating psychiatric diagnosis in biological terms \cite{miller2010beyond} will require a systematic exploration of the low-level mechanisms giving rise to the behavioural manifestations of mental disorders \cite{deco2014great, murray2018biophysical}. We expect that many of the ideas and methods here proposed will seamlessly translate into the field of computational psychiatry, even for the study of disorders which do not include altered consciousness as a defining feature (e.g. depression).

In the same way that scientific inquiry has eventually succeeded explaining seemingly mysterious phenomena such as heat (in terms of kinetic considerations), combustion (in terms of chemical reactions) and genes (in terms of molecular replication), it is reasonable to expect that consciousness will also be explainable someday in mechanistic terms. If this is to happen, the perspective of bottom-up modelling is likely to play a crucial role, as it was the case for the three aforementioned examples. It is our hope that the present proposal will serve both as an encouragement and as a roadmap to invest future research efforts in the computational modelling of altered states of consciousness.

\vspace{6pt}

\authorcontributions{Conceptualization, R.C.,  R.H., P.A.M.M, F.E.R, Y.S.P and E.T; methodology, R.C.,  R.H., P.A.M.M, F.E.R, Y.S.P and E.T; writing--original draft preparation, R.C.,  R.H., P.A.M.M, F.E.R, J.P, Y.S.P and E.T; writing--review and editing R.C.,  R.H., P.A.M.M, F.E.R, J.P, Y.S.P, and E.T} 

\funding{R.C. was supported by Fondecyt Iniciaci\'{o}n 2018 Proyecto 11181072. R.H. was funded by CONICYT scholarship CONICYT-PFCHA/Doctorado
Nacional/2018-21180428. P.M. was funded by the Wellcome Trust (grant no.
210920/Z/18/Z). F.R. was supported by the Ad Astra Chandaria Foundation. E.T. and Y.S.P. were supported by ANPCyT (Argentina), grant PICT-2018-03103}


\conflictsofinterest{The authors declare no conflict of interest.} 

\abbreviations{The following abbreviations are used in this manuscript:\\

\noindent 
\begin{tabular}{@{}ll}

NCC & Neural correlates of consciousness\\
DMF & Dynamic mean field\\
fMRI & Functional magnetic resonance imaging\\
BOLD & Blood oxygen level–dependent\\
PET & Positron emission tomography\\
DTI & Diffusion tensor imaging\\
EEG & Electroencephalography\\
MEG & Magnetoencephalography\\
IIT & Integrated Information Theory\\
GNW & Global neuronal workspace\\
EBH & Entropic brain hypothesis\\
LZc & Lempel-Ziv complexity\\
FCD & Functional connectivity dynamics \\
PCI & Perturbational complexity index\\
TMS & Transcranial magnetic stimulation\\
PILI & Perturbative Integration Latency Index\\
LSD & Lysergic acid diethylamide\\
AAL & Automated anatomical labelling\\
DOC & Disorder of consciousness\\
GABA & Gamma-aminobutyric acid \\
RSN & Resting-state networks \\

\end{tabular}}

\reftitle{References}
\externalbibliography{yes}
\bibliography{biblio.bib}

\end{document}